# On-chip directional octave-spanning supercontinuum generation from high order mode in near ultraviolet to infrared spectrum using AlN waveguides

Hong Chen[1], Jingan Zhou[1], Dongying Li[1], Dongyu Chen[2], Abhinav K. Vinod[3], Houqiang Fu[1], Xuanqi Huang[1], Tsung-Han Yang[1], Jossue A. Montes[1], Kai Fu[1], Chen Yang[1], Cun-Zheng Ning[1,4], Chee Wei Wong[3], Andrea M. Armani[2], and Yuji Zhao[1,*]

[1]School of Electrical, Computer and Energy Engineering, Arizona State University, Tempe, AZ 85287, USA
[2]Mork Family Department of Chemical Engineering and Materials Science, University of Southern California, Los Angeles, CA 90089, USA
[3]Department of Electrical Engineering, University of California, Los Angeles, CA 90095, USA
[4]Department of Electronic Engineering, Tsinghua University, Beijing 100084, China
[*]Correspondence to: yuji.zhao@asu.edu

**Abstract:** On-chip ultraviolet to infrared (UV–IR) spectrum frequency metrology is of crucial importance as a characterization tool for fundamental studies on quantum physics, chemistry, and biology. Due to the strong material dispersion, traditional techniques fail to demonstrate the device that can be applied to generate coherent broadband spectrum that covers the full UV–IR wavelengths. In this work, we explore several novel techniques for supercontinuum generation covering near-UV to near-IR spectrum using AlN micro-photonic waveguides, which is essential for frequency metrology applications: First, to create anomalous dispersion, high order mode ($TE_{10}$) was adopted, together with its carefully designed high efficiency excitation strategies. Second, the spectrum was broadened by soliton fission through third order dispersion and second harmonic generation, by which directional energy transfer from near-IR to near-UV can be obtained. Finally, high quality single crystalline AlN material was used to provide broadband transparency from UV to IR. Under decently low pulse energy of 0.36 nJ, the experimental spectrum from supercontinuum generation covers from 490 nm to over 1100 nm, with a second harmonic generation band covering from 405 nm to 425 nm. This work paves the way towards UV–IR full spectrum on-chip frequency metrology applications.

On-chip supercontinuum generation spanning from UV through IR with low (sub-nJ) powers has been a quest of researchers since the original demonstration in the 1960's [1,2]. An integrated system would enable research efforts and technology in white-light emitters [3,4], ultrahigh-resolution spectroscopy [5,6], high-speed interconnection [7], and quantum states generation [8,9]. While attempts have been made using a wide range of material systems [10-20] and nonlinear phenomena, this achievement has eluded the research community due to fundamental principles. Namely, the majority of attempts relied on a combination of self-phase modulation (SPM) and soliton fission (SF) processes where the solitons were perturbed by 4[th] order dispersion in fundamental transverse electric (TE) / transverse magnetic (TM) modes. Moreover, solitons perturbed by 4[th] order dispersion emits dispersive waves (DWs) in both blue and red shifting spectral, which degrades the energy efficiency for short wavelength applications. Despite the success in achieving broadband spectrum this strategy failed to reach below 500nm due to limits in material crystallinity and reliance on long wavelength pump sources. One exception is work based on silica ridge waveguides which successfully reached into the UV range [13]. However, the low nonlinear coefficient of silica required large pump energies to be used. Therefore, to overcome this limit, a higher nonlinear material must be used. One such material is AlN. In a preliminary study, the high $\chi^{(2)}$ of AlN was used to achieve directional nonlinear energy transfer [18]. This approach enabled a spectrum covering over 100THz within the UV region using only 0.237nJ. Unfortunately, this system was limited to operation in the normal GVD range. In this work, these limitations are overcome to achieve supercontinuum generation from near-UR to near-IR with sub-nJ pulse energy.

To achieve supercontinuum generation covering near-UV to near-IR spectrum, several nonlinear optical processes can be proposed as the potential broadening mechanisms for the device design with both advantages and limitations. For example, SPM is the most commonly observed process in the initial stage of spectrum broadening, however it requires excessive high excitation power to further expand the spectrum [21]; Four-wave mixing is applied in numerous studies [22] of micro-comb systems, however it generally requires fine resonating device structures; Harmonic generation provides good directional energy transfer, however it has limited bandwidth due to the phase-matching conditions [18]; Self-steepening has the potential for broadband directional supercontinuum generation in UV–visible range, however it requires small and flat dispersion near pumping wavelengths to build the steep temporal structures [15], which is hindered by the strong material dispersion in the near-visible spectrum; Other nonlinear processes such as Stokes scattering [23] and modulation instability [23] are usually accompanied with strong phase noise thus are not favorable. To tackle the issues mentioned above, several advances in engineering and in science are simultaneously combined in an AlN integrated waveguide [Figure 1(a)]. First, to create anomalous dispersion, a high order waveguide mode ($TE_{10}$) was used. To efficiently excite this mode, a high efficiency excitation strategy was designed and fabricated. Second, the spectrum was broadened by soliton fission through third order dispersion and second harmonic generation. This approach allowed directional energy transfer from near-IR to near-UV to be obtained. Third, high quality single crystalline AlN material was used to provide broadband transparency from UV to IR. The development and use of high quality single crystalline of AlN improved the optical performance of the device, allowing the nonlinear processes to occur simultaneously and with high efficiency. As a result of these advances, supercontinuum generation from 490 nm to over 1100 nm, with a second harmonic generation band covering from 405 nm to 425 nm is achieved with only 0.36nJ of pulse energy. This work paves the way towards UV–IR full spectrum on-chip frequency metrology applications.

**Results**

To optimize the dispersion of the AlN waveguide, the modal dispersion for a series of waveguide geometries was computed from an in-house made numerical solver based on finite-difference method (FDM) provided in [24]. The computed dispersions were compared with commercial software (Lumerical modes) to confirm accuracy. These calculations were used to direct the device design and mode selection. For example, the calculated GVDs in a 1.1 μm × 1.2 μm (W×H) waveguide is shown in Fig. 1(b). Due to the highly dispersive material property, $TE_{00}$ and $TM_{00}$ modes exhibit normal dispersion below 1000 nm, while $TE_{10}$ mode shows anomalous dispersion above 740 nm. Anomalous dispersion supports the formation of solitons near the operating wavelengths of Ti:sapphire laser. The zero dispersion wavelength (ZDW) of this $TE_{10}$ mode is widely tunable by varying waveguide widths, as illustrated in Fig. 1(c).

Using these calculations as the basis for device fabrication design, single crystalline AlN thin films were epitaxially grown by metalorganic chemical vapor deposition (MOCVD) on sapphire substrates. The surface roughness was determined by atomic force microscopy (AFM) in a 5 μm × 5 μm scan with a root mean squared (RMS) roughness of ~ 3 nm. Refractive indexes of the thin films were measured by ellipsometry. Figure 1(d) shows the schematic view of the device structure of the AlN waveguide, where the waveguide geometrical parameters, such as width (W), height (H), and sidewall angles (θ), are defined. Waveguides were fabricated from 1.2 μm thick films (H) with widths ranging from 0.8 μm to 1.6 μm and lengths around 0.6 cm. Figure 1(e) shows the representative cross-section transmission electron microscopy (TEM) image of the AlN waveguide. The crystalline nature of the AlN thin film is clearly evident. Additional material data and waveguide fabrication details are in the Methods section.

The experimental setup is depicted in Fig. 1(f). A standard Ti:sapphire laser (Spectra Physics) with 100 fs pulse width (full-width half maximum) and 82 MHz repetition rate was used. To couple light into the on-chip waveguides from the free-space laser, end-fire coupling was used. The out scattered light from the waveguides was collected by a linear CMOS camera (Thorlabs DCC1240) and used to perform propagation loss estimation. A tapered fiber was placed near the output port of waveguides and the output signal was fed into optical spectrum analyzer (Yokogawa AQ6373B). Coupling strategies to efficiently excite the high order modes are discussed in the Methods section.

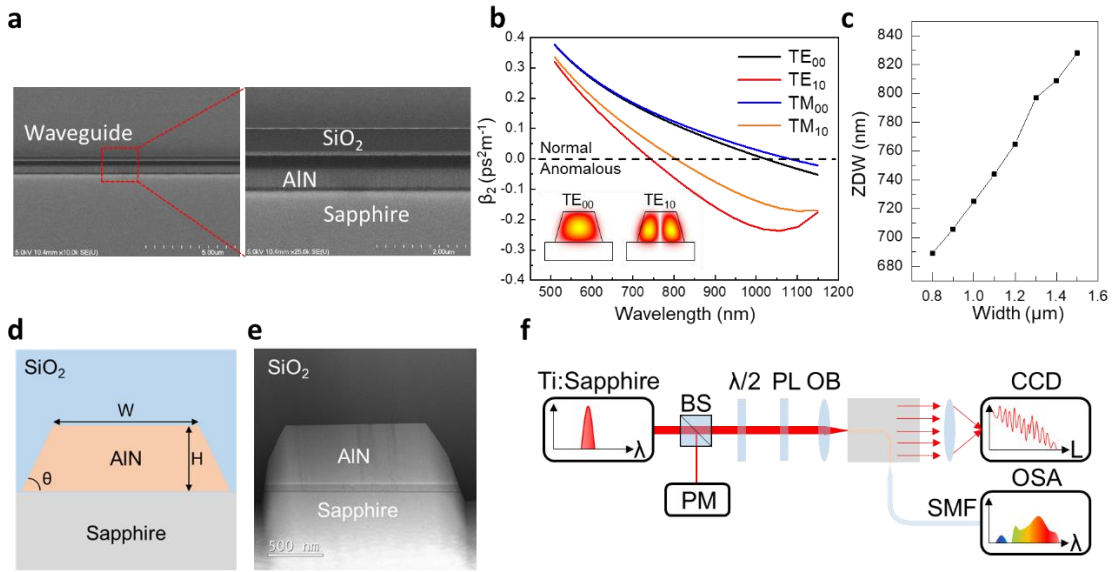

Figure 1. (a) Scanning electron microscope image of AlN waveguides fabricated in this research. The $SiO_2$ hardmask is remaining on the top. Smooth sidewall can be identified from the zoomed image. (b) Calculated group velocity dispersions of $TE_{00}$, $TE_{10}$, $TM_{00}$, and $TM_{10}$ modes in a 1.1 μm × 1.2 μm (W × H) AlN waveguide, the insertion shows the mode profiles ($|E_x|$) of $TE_{00}$ and $TE_{10}$ modes, which are the modes that utilized within this work. (c) Zero-dispersion

wavelength vs. waveguide geometry, the height was kept at 1.2 μm. (d) Schematic view of the device structure of the AlN waveguide, where the waveguide geometries such as width (W), height (H), and sidewall angles (θ) are defined. (e) Transmission electron microscope image of a typical AlN waveguide fabricated in this work. (f) Experimental setup implemented in this work.

A broadband supercontinuum spectrum is shown in Fig. 2(a). The main spectrum spans from 490 nm to 1100 nm, which is over one octave. Due to the large second order susceptibility of AlN, a secondary spectrum was observed from 405 nm to 425 nm. The pumping wavelength was slightly tuned near 800 nm, and the best spectrum [Figure 2(a)] was recorded at $\lambda = 810$ nm. Minor changes were observed when varying the pumping wavelength from 800 nm to 820 nm.

The broadened spectrum can be identified as the composition of $TE_{00}$ and $TE_{10}$ modes. Due to the ultra-short pulse width and different group velocities, the $TE_{00}$ and $TE_{10}$ modes split in the temporal domain after a propagation length of a few microns. Therefore, throughout this study, the mutual nonlinear coupling between $TE_{00}$ and $TE_{10}$ modes were neglected. By comparing the propagation loss and on-chip average power of $TE_{00}$/$TE_{10}$ modes, the power difference between $TE_{00}$ and $TE_{10}$ mode can be justified to be ~ 10 dB. According to the numerical simulation, the excitation efficiencies of $TE_{00}$ and $TE_{10}$ modes are 15% and 20%, respectively (see S.I.). On-chip average power was estimated to be ~ 50 mW, thus the average power within $TE_{10}$ mode was estimated to be 30 mW, corresponding to a pulse energy of 0.36 nJ, which is low when compared with other supercontinuum generation methods at short wavelengths [13,15,17].

By solving the nonlinear Schrodinger's equation (NSE), the simulated spectrum for each mode can be obtained which is also shown in the spectrum plot in Fig. 2(a). Details in regard to NSE simulation can be found in "method". As predicted in Figure 1(c), the $TE_{00}$ mode was propagating within normal dispersion region, therefore self-phase modulation is the major contributor to its broadening mechanism. In contrast, for the $TE_{10}$ mode, the pulse was split into its continent fundamental solitons, and the solitons were perturbed by TOD. Therefore, DWs were emitted near 500 nm through non-solitonic radiation, enabling the formation of the broadband supercontinuum in the $TE_{10}$ mode case.

Under high power excitations, complex physical processes are expected such as strong thermal effects [25], multi-photon absorption [26], free carrier dispersion [26], cross-phase modulation [27], modulation instability [28], and mode avoid-crossing [29]. Under such conditions, the NSE only provides good estimation on spectrum width, but cannot accurately predicts the spectrum shape at all wavelengths. Therefore, to further confirm the non-solitonic radiation process, a series of low excitation power measurements were performed, and the corresponding spectra were recorded (Figs. 2(b)-2(d)). When the average pumping power is below 10 mW, the self-phase modulation process is responsible for the symmetric broadening as indicated in Figs. 2(b) and 2(c). When the average pumping power reaches 10 mW which corresponds to a pulse energy of only 0.12 nJ, significant asymmetric broadening starts to initiate as shown in Fig. 2(d). The development of supercontinuum spectrum with increasing power can be further illustrated using Fig. 2(e). The spectral broadening was first initiated symmetrically by SPM (point 1 to 2), and then asymmetrically expanded by DW emission (point 2 to 3). From point 3 to 4, blue shifts of DW can be observed, which is a result from the power dependent phase matching condition for DWs [23]:

$$\beta(\omega) = \beta(\omega_s) + \beta_1(\omega - \omega_s) + \frac{1}{2}\gamma P_s \qquad (1)$$

where $\beta$ indicates wavevector. $\omega_s$ and $\omega$ indicate angular frequency of soliton and DW, respectively. $\beta_1$ is equal to $\partial\beta/\partial\omega$. $P$ is the power within soliton modes. $\gamma$ is the nonlinear parameter that described by $(\omega n_2)/(cA_\text{eff})$, where the $c$ and $A_\text{eff}$ indicate vacuum speed of light and effective modal area, respectively. The blue shifts of DW with increasing power can be attributed to the power dependent term in the Eq. (1).

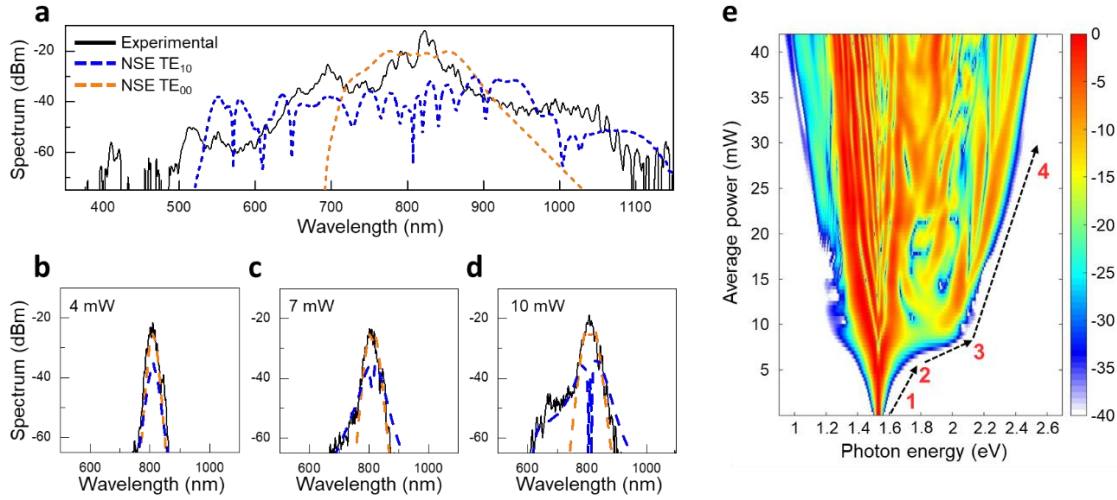

Figure 2. (a) Experimental results of the supercontinuum spectrum (black solid), and the nonlinear Schrodinger's equation (NSE) simulated spectrum of $TE_{00}$ (orange dash) and $TE_{10}$ (Blue dash) modes for the AlN waveguide. The spectrum is generated with a pump wavelength of 810nm and at average pump power of 30 mW. (b)-(d) The experimental (black solid) and the NSF simulated results (orange and blue dash) under low power pumping conditions. (e) Spectral evolutions of the AlN waveguide at different average pumping powers (within $TE_{10}$ mode) used in the measurements. From point 1–2, 2–3, and 3–4, the broadening mechanism was governed by SPM, SF, and optical power dependent phase-matching between soliton and DW.

Given the dependence shown in Figure 2e, it is apparent that, unlike the prior results on supercontinuum generation that relies on the 4[th] order dispersion, the spectrum obtained in this work exhibits strong dependence on waveguide geometry due to the near ZDW pumping. To further investigate this dependence and create a generalizable waveguide design strategy, the GVD, TOD, and 4th order dispersion curves for AlN waveguides with different widths are calculated (Figure 3(a)). A waveguide width below 1200 nm is required to support the solitons near pumping wavelengths. It is also noteworthy that below a waveguide width of 800 nm, perturbation on solitons are mainly due to the 4[th] order dispersion from the sudden decrease of $\beta_3$, which results in DWs emission towards unwanted spectral wavelengths. Moreover, as is estimated in Fig. 3(b), the propagation loss of the $TE_{10}$ mode increases exponentially with reducing waveguide widths, which exceeds 50 dB/cm at a waveguide width of 1000 nm. This strong loss prohibits the formation of fundamental solitons [26], which in turn prevents the DW emission.

Combining the observations in Figs. 3(a) and 3(b), the design space of the AlN waveguide devices for this work is limited within a waveguide width from ~ 1000 nm to ~ 1200 nm. To verify these calculations, devices outside of this optimum range are fabricated and characterized. Figure 3(c) illustrates the measured spectrum and the NSE simulated pulse spectrum for the $TE_{00}$ mode in AlN waveguides with widths of 800 nm and 1400 nm. The broadening processes in these two waveguide designs can be clearly identified as SPM, and the average pumping power within the $TE_{10}$ was estimated to be ~ 20 mW, which is two times higher than that in case of Fig. 2(d).

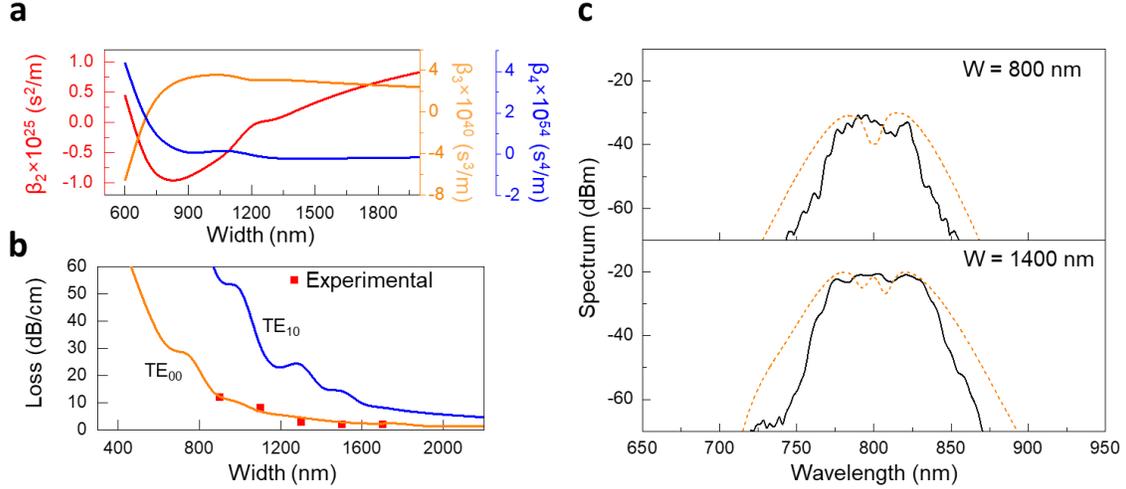

Figure 3. (a) The calculated GVD ($\beta_2$), TOD ($\beta_3$), and 4$^{th}$ order dispersion ($\beta_4$) curves for AlN waveguides with different widths, represented by red, orange, and blue curves, respectively. (b) Experimentally measured propagation loss (square points) in dB/cm and the calculated propagation loss for $TE_{00}$ (orange curve), $TE_{10}$ (blue curve) modes for AlN waveguides with different widths. (c) Measured spectrum (black curve) and the NSE simulated pulse spectrum (orange dash curve) for $TE_{00}$ mode in AlN waveguides with widths of 800 nm and 1400 nm. The spectrum was mainly broadened by SPM of the fundamental $TE_{00}$ mode, therefore the simulated spectral for the $TE_{10}$ mode are not shown here.

While the generation of the primary spectrum from 490 nm to 1100 nm with only 0.36nJ is remarkable, the simultaneous generation of an even higher energy (lower wavelength) secondary spectrum spanning from 405 to 425 nm further increasing the potential impact of this system. To understand the physical mechanism that gives rise to this secondary spectrum, we solve for the mode dispersion near the fundamental and the SHG wavelengths, and the mode profiles of high order modes near SHG wavelengths are depicted in Fig. 4(a). Figure 4(b) shows the modal dispersion of $TE_{00}$ mode and $TE_{10}$ mode near 800 nm, and the dispersion of several high order modes near 400 nm. By optimizing the waveguide coupling design, modal phase-matching was achieved at 398 nm and 412 nm. Furthermore, Fig. 4(b) also shows the dispersion curve of $TE_{10}$ mode in the near-visible region. Since the phase-matching point is far away from 400 nm, and the two phase-matching points have wavelength difference of over 40 nm, we neglect the SHG effect from the $TE_{10}$ mode. The experimental identified second harmonic signal was recorded at 407 nm and 421 nm and is shown in Fig. 4(c). The MOCVD growth and the dry etching processes for the AlN waveguides may cause minor geometry discrepancy between the simulation and the experiment, which resulted in the small differences between the simulated and experimentally recorded phase matching points. Since the phase-matching wavelengths are dependent on the device geometry, the location of SHG signal is invariant to pumping wavelengths, which is evident in Fig. 4(c).

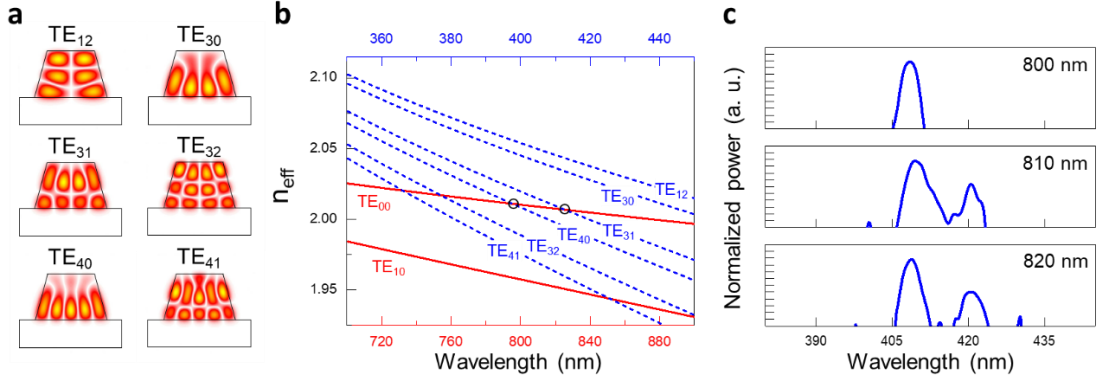

Figure 4. (a) Mode profile ($|E_x|$) of each high order mode computed at 400 nm wavelength. (b) The modal dispersion of $TE_{00}$ mode and $TE_{10}$ mode near 800 nm (red curves) and the dispersion of several high order modes near 400 nm (blue dash curves) for the secondary spectrum from 405 to 425 nm in Fig. 2(a). Phase matching wavelengths can be determined at the crossing points of the curves. The two peaks measured from experiment can be identified at the two circle-marked points. (c) SHG signal near 400 nm at different pumping wavelengths, the locations of the SHG peaks were invariant to pumping wavelengths.

The optical coherence of the spectrum can be determined by applying the first-order coherence function [10,14,17]:

$$\left|g_{12}^{(1)}(\lambda)\right| = \left|\frac{\langle E_1^*(\lambda)E_2(\lambda)\rangle}{\sqrt{\langle|E_1(\lambda)|^2\rangle\langle|E_2(\lambda)|^2\rangle}}\right| \quad (2)$$

which is the ensemble average of multiple supercontinuum pulses. The simulation took 100 supercontinuum pulses with a standard shot noise at the input spectra, and the obtained first-order coherence of $TE_{10}$ supercontinuum is shown in Fig. 5. The coherence is close to unity from visible to near-infrared, which indicates the broadening strategy applied in this work is robust to noise.

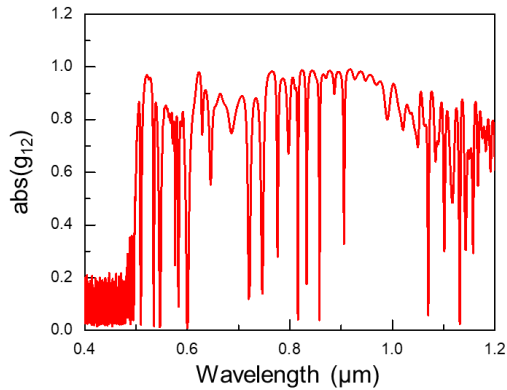

Figure 5. The simulated first-order coherence function vs. wavelengths for AlN waveguide with a width of 1.2 μm. The coherent function is near unity across the whole spectrum, indicating that the broadening mechanism is robust to shot noise.

**Conclusion**

In conclusion, we obtained supercontinuum generation from dispersion engineered AlN waveguide, the optical energy of which directionally transferred from near-IR towards near-UV wavelengths. The main spectrum covers from 490 nm to over 1100 nm with a secondary SHG spectrum that covers from 405 nm to 425 nm. Near-visible pumping was implemented to improve energy efficiency. To overcome the strong material dispersion, the AlN waveguides were designed to support high order modes. The experiment results were compared with the theoretical results from NSE simulation, which confirmed that DWs were generated from solitons perturbed by TOD, allowing for directional energy transfer. Further investigations on high order dispersion terms and propagation losses reveal that the SF process can only be initiated within a narrow window of waveguide width, which was supported by experimental observations. The SHG spectrum was investigated by solving modal dispersions near 800 nm and 400 nm, and the corresponding phase-matching wavelengths were determined. Simulation on first-order coherence function suggests that the spectrum broadening procedure is robust to noise, which is essential for frequency metrology related applications. This work provides novel waveguide design strategies and paves the way towards UV–IR frequency comb generations.

**Methods:**

**Fabrication AlN waveguides.** The AlN thin films were coated with a ~ 600 nm $SiO_2$ layer using plasma enhanced chemical vapor deposition (PECVD), followed by an 80 nm Cr layer using electron-beam evaporation. The two layers served as hardmasks for the dry etching processes. Photoresist ma-N 2403 was used to perform electron-beam lithography (EBL). The Cr layer was etched away using a user developed (chlorine + argon) reactive ion etching (RIE) process, while the $SiO_2$ layer was removed by standard anisotropic RIE etching process developed by ASU NanoFab. The AlN waveguide patterns were defined using a user developed inductively coupled plasma (ICP) etching with $Cl_2$, $BrCl_3$, and Ar chemistries at a bias voltage near 300V. The waveguides were coated by 2 μm $SiO_2$ coating layers to reduce scattering loss. After fabrication processes, samples were cut and polished down to 0.1 μm grade. The fabricated waveguide has ~ 50 to 100 nm width variance due to the underneath isotropic etching during Cr etching process, this variance was involved in numerical simulations. Detailed process flow and scanning electron microscope (SEM) images of the fabricated AlN waveguides can be found in S.I.

**Simulation on pulse propagation.** The pulse propagation along the waveguide was simulated using an in-house developed solver for nonlinear Schrodinger's equation using split-step Fourier method [23], where the 4[th] order Runge-Kutta method [30] was also implemented to reduce computation load. The format of the nonlinear Schrodinger's equation was:

$$\frac{\partial A}{\partial z} = -\frac{\alpha}{2}A + i\sum_{k\geq 2} i^k \frac{\beta_k}{k!}\frac{\partial^k A}{\partial t^k} + i\gamma[|A|^2 A + \frac{i}{\omega_0}\frac{\partial}{\partial t}(|A|^2 A) - T_R A\frac{\partial |A|^2}{\partial t}] \quad (3)$$

where $A(z,t)$ denotes the pulse slow varying amplitude, $\alpha$ is the propagation loss, $\beta_k$ is the kth order dispersion, $\gamma$ is the nonlinear parameter, and $T_R$ is the Raman response parameter. It should be noted that we keep the most simplified Raman response form in Eq. (3), and the $T_R$ was set to zero as no significant red-shifted spectrum was observed through this whole study. Dispersion terms were truncated at $6^{th}$ order, but it's noteworthy that only TOD plays significant role in modifying the spectrum. The Kerr refractive index of AlN was estimated to be $2.5 \times 10^{-19}$ m$^2$/W, which gives a nonlinear parameter $\gamma \sim 3$ W$^{-1}$m$^{-1}$ depending on the mode effective area. The initial pulse shape was assumed to be Gaussian type with a full-width-at-half-maximum (FWHM) of 100 fs.

**High order mode excitations.** We adopted different waveguide excitation strategies in this work, the detailed descriptions (Pros. and Cons.) of each excitation methods are discussed in S.I.. The most efficient excitation method was applying normal taper with a $\sim 5$ μm taper width. Under low power operation, the beam was first focused at the center of taper, then slowly varying the nano-stage to move the focal point from center to the edge of taper facet. The out-scattered light was monitored by CMOS camera at the same time. Since the TE$_{10}$ mode has higher scattering loss comparing to the TE$_{00}$ mode, when the out-scattered light reaches its maximum (as recorded by CMOS camera), the excitation efficiency of TE$_{10}$ mode will be near its maximum. At this stage, the location of focusing point is near the edge of taper facet, which can be confirmed by the CMOS camera. An alternative way to excite TE$_{10}$ mode was to operate the system under higher power (near 30 mW on-chip average pumping power). By slowly varying the focal point from center to the edge of taper, DWs near 600 nm were generated due to the increasing excitation efficiency of the TE$_{10}$ mode. The red radiation can be directly observed under microscope, and the excitation efficiency of the TE$_{10}$ mode can be optimized by maximizing the red radiation.

**Estimation of propagation loss.** The propagation losses of TE$_{00}$ modes were experimentally characterized by collecting out-scattered light along the propagation direction using a CMOS camera similar to our previous investigation in [31]. The typical scattered optical power versus propagation length is provided in S.I.. By adopting semi-log plotting, the decay slope is proportional to propagation loss of TE$_{00}$ mode in dB/cm. The low loss of TE$_{00}$ mode was further confirmed by measuring its intrinsic quality factor of ring resonators (See S.I.).

The losses of TE$_{00}$ modes with different waveguide widths were fitted by an in-house made dyadic Green's function (DGF) solver to obtain the sidewall roughness. The detailed description on DGF can be found in S.I.. DGF fitting gives sidewall roughness around 4 nm, which agrees well with standard optimized ICP etching process. By plugging in TE$_{10}$ mode profile and sidewall roughness, the scattering loss can be estimated.

**Supplementary information**

The supplementary information includes material characterization, fabrication process flow, waveguide performance evaluation, high order mode excitation efficiency, propagation loss estimation, and auxiliary experimental data.

**Acknowledgments**

This work was supported by the grant from Army Research Office monitored by Dr. Michael Gerhold. We gratefully acknowledge the use of facilities within the LeRoy Eyring Center for Solid State Science, ASU NanoFab, and ASU CLAS ultra-fast laser facility. We received great supports from ASU nanophotonics group lead by Prof. Cun-Zheng Ning. The authors would also like to thank Dr. Su Lin and Dr. Douglas Daniel for their great assistance in the early state of this research.

**Author contributions**

Y.Z. initiated this research and supervised the project progress. H.C. coded the numerical solvers, designed the waveguides, fabricated the devices, and performed the optical testing. J.Z. and D.L. (supervised by C.Z.N.) participated in the optical testing of the late devices within the laboratory. D.C. (supervised by A.M.A.) and A.K.V. (supervised by C.W.W) evaluated waveguide performance of GaN and AlN by measuring the Q factor of the resonators and provided valuable discussions. H.C., J.Z., H.F., X.H., T.H.Y., and J.A.M. carried out the material characterizations (ellipsometry, XRD, AFM, SEM, etc.). K.F. and C.Y. assisted the device fabrication. H.C. and Y.Z. wrote the manuscript with inputs and suggestions from other authors. All authors participated in data analysis and discussions, and contributed significantly to this work.